\documentclass
{mn2e}
\usepackage{epsfig}
\usepackage{amsmath, amssymb,bm}

\def \be{\begin{equation}}
\def \ee{\end{equation}}

\def \msun{\rm M_{\odot}}

\def \me{{\dot M_{\rm Edd}}}
\def  \le{{L_{\rm Edd}}}

\begin{document}
\title[Joining the Dots: High Redshift Black Holes] {Joining the Dots:  High Redshift Black Holes}

\author[Andrew King] 
{\parbox{5in}{Andrew King$^{1, 2, 3}$ 
}
\vspace{0.1in} \\ $^1$ School of Physics \& Astronomy, University
of Leicester, Leicester LE1 7RH UK\\ 
$^2$ Astronomical Institute Anton Pannekoek, University of Amsterdam, Science Park 904, NL-1098 XH Amsterdam, The Netherlands \\
$^{3}$ Leiden Observatory, Leiden University, Niels Bohrweg 2, NL-2333 CA Leiden, The Netherlands}

\maketitle

\begin{abstract}
A recent paper (King, 2024) suggested that emission from the central supermassive black holes in high--redshift galaxies must be tightly collimated by the effects of partly
expelling a  super--Eddington mass supply. I show here that this idea predicts that these galaxies should produce very little detectable rest--frame X--ray emission, appear Compton thick, and show no easily detectable sign of outflows. All of these properties agree with current observations. To produce these effects, the mass supply to the black holes should exceed the Eddington rate by factors $\sim 50 - 100$, which appears in line with conditions during the early growth of the holes. I note that theoretical derivations of the ratio of black hole mass to host galaxy stellar mass already predict that this should increase significantly at high redshift, in line with recent observations.
\end{abstract}


\footnotetext[1]{E-mail: ark@astro.le.ac.uk}
\section{Introduction}
\label{intro}

Recent JWST observations of galaxies at redshifts $z \gtrsim 4$ have stimulated intense interest. These results open a path to understanding the genesis of the central supermassive black holes (SMBHs) astronomers now generally believe lie at the centre of every galaxy. 
Probably the most puzzling feature one has to explain has been the finding of very high SMBH masses $M \gtrsim 3\times 10^9\msun$ (cf e.g. Willott, McLure \& Jarvis, 2003) at redshifts $z \sim 6$: this is so early in cosmic time that these masses could not have grown via conventional accretion on to initially stellar--mass seeds. 

In an earlier paper (King, 2024) I suggested a possible escape from this difficulty, by appealing to the analogy with the ultraluminous X--ray sources (ULXs) found in nearby galaxies. These are now known to be stellar--mass binary systems where the compact accretor (a black hole or neutron star) is fed mass by its companion at rates far exceeding that which would produce the Eddington luminosity $L_{\rm Edd}$ (see King, Lasota \& Middleton, 2023, for a review). 

This super--Eddington mass supply has two effects: most of it is blown away as a 
dense high--speed wind, and a significant fraction of the accretion luminosity -- which is of order $L_{\rm Edd}$ -- produced by the remainder emerges in a strongly collimated (`beamed') component escaping down narrow channels through this wind, along the rotational axis of the central accretion disc\footnote{In this paper `beaming' simply means collimation by scattering, and not more exotic processes such as relativistic beaming.}. (The disc axis near its centre is probably always aligned to the black hole spin direction through warping by the Lense--Thirring effect, whatever the orientation of the disc far from the hole. See Section 5.3 of King, 2023 for a discussion.) The very high specific intensity of this component means that observers lying in one of the beams and assuming the emission was spherically symmetrical would assign a very high luminosity -- and so accretor mass -- to the source. In the early years after their discovery this led to claims that ULXs contained black holes with `intermediate' masses $\sim 100 - 10^4\msun$. 
Recent observations have now shown conclusively that they are instead normal stellar--mass X--ray binaries passing through a phase of very rapid mass transfer. 

Extending these ideas to SMBH fed at super--Eddington rates provides a simple reason why emission from the close vicinity of the black hole might suggest a very high apparent (i.e. assumed spherical) luminosity, and so a much higher mass than
the true one. Simultaneously, the presence of high--speed outflows means that 
one cannot safely deduce the SMBH mass by using virial indicators which assume bound orbiting gas. It seems worthwhile to ask if these considerations apply quite generally to all SMBH at high redshift, and perhaps give some insight into related questions. 



\section{Tests of Beaming}
The idea of beaming for the black holes of high--redshift AGN is open to straightforward test. If 
the beaming fractions $b$ of the high--$z$ AGN (the ratio of the
true emitted luminosity to the value found by assuming spherical symmetry) are small,
as for example required to reduce the estimated SMBH masses, the beams are narrow. 
Then randomly--placed observers will usually be situated outside them,
%
and so almost all high--$z$ galaxies should show very little emission originating
very close to the central SMBH, that is, almost no detectable rest--frame X--ray emission. 

This prediction is in clear agreement with the deep Chandra observations of  Maiolino et al., (2024) (henceforth M24), where only 2 out of 71 galaxies are detected in rest--frame X--rays. We can get a simple estimate of the physical conditions required for this, by using the standard formula giving $b$, i.e.
\be
b \simeq \frac{73}{\dot m^2} 
\label{beam}
\ee
(King, 2009) where $\dot m = \dot M/\me \gg 1$ is the ratio of the mass supply to the value giving the Eddington luminosity. This formula was developed for ULXs, which are stellar--mass binaries with mass transfer rates far above the Eddington rate, but we can extend its use to SMBH systems, as black--hole accretion is a scale--free process. Simple physical arguments (King, 2009)
show that $b \propto \dot m^{-2}$, and for the ULXs this implies a striking inverse relation $L \propto T^{-4}$ between soft X--ray luminosity $L$ and its blackbody temperature $T$ for a large sample of systems. This relation is observed (Kajava \& Poutanen, 2009), and sets the normalization factor 73 in eqn (\ref{beam}). 

If we assume that all of the M24 sample have similar beaming fractions, this would suggest $\langle{b}\rangle \sim 2/71$ and so
\be
\langle\dot m\rangle \sim 50 
\label{mdot}
\ee
for this sample. 
The super--Eddington mass outflow rate automatically implies that the winds carrying off the ejected gas are Compton thick (see the Appendix) which again agrees with observation (M24). This additionally means that the outflows themselves must be effectively unobservable, as M24 find.

Recent observations of the so--called `Little Red Dots' (henceforth LRDs) give further insight into the physics of galaxies born early in cosmic time. Pacucci and Narayan (2024) review LRD properties in some detail: they are very red, compact (effective radii $r_e \sim 150$~pc, with some even smaller, i.e. $r_e< 35$~pc). LRDs are $\sim 10 -100$ times more numerous than the faint end of the quasar luminosity function. Their very small $r_e$ values imply extreme stellar densities, so that stellar collisions should be very efficient in feeding a central black hole (Guia, Pacucci \& Kocevski, 2024) if one is present. Some of these stars may accrete quietly to a central SMBH as they do not fill their tidal lobes before crossing the event horizon. But some must be tidally disrupted and produce X--ray emission if the black hole mass $\lesssim 10^7\msun$. 

As with the  M24 sample, LRDs are intrinsically weak in X--rays. For example, in a comprehensive sample of 341 LRDs spanning several JWST fields, Kocevski et al. (2024) found only two X--ray
detected LRDs, despite that fact that they show signs usually associated with active SMBH, such as broad emission lines (Greene et al., 2024). The same argument used to give (\ref{mdot}) would give an averaged value
\be
\langle\dot m\rangle \sim 116
\label{mdot2}
\ee
for the LRD sample.

\section{Discussion}

The relatively tight beaming discussed in this Letter predicts that searches for
rest--frame X--ray emission from high--redshift galaxies should produce extremely
sparse results. The results of two separate deep surveys with different samples agree closely with this prediction, and with the additional properties of being Compton thick and showing no detectable sign of outflows. The physical conditions required for these outcomes are that the host galaxies should supply mass to the SMBH involved at with an Eddington factor $\dot m \sim 50 - 100$.  This is not a particularly stringent requirement: in LRDs the very high stellar densities seem likely to induce mass infall at rates close to the dynamical value 
\be
\dot M_{\rm dyn} \sim \frac{f_g\sigma^3}{G},
\ee
 where $\sigma$ is the central velocity dispersion and $f_g$ the gas fraction. To see this I note that virial equilibrium in a galaxy can support a gas mass $M_g \sim \sigma^2 f_g R/G$ at radial distance $R$. If this is destabilized (at high redshift, perhaps by interaction with a nearby galaxy) it must fall towards the central SMBH on a dynamical timescale $t_{\rm dyn} \sim R/\sigma$. This gives a mass supply rate 
 $\dot M \sim M_g/t_{\rm dyn} \sim \dot M_{\rm dyn}$ for a dynamical time 
 $R/\sigma$, which is of order $5\times 10^6$~yr for $R \sim 1$~kpc and 
 $\sigma \sim 100~{\rm km\, s^{-1}}$.
For SMBH close to the $M - \sigma$ relation the dynamical rate gives an Eddington factor
\be
\dot m = \frac{\dot M_{\rm dyn}}{\dot M_{\rm Edd}} \simeq 
\frac{64}{\sigma_{200}} \simeq \frac{54}{M_8^{1/4}},
\ee
where $\sigma_{200} = \sigma/200~{\rm km\,s}^{-1}$ and $M_8 = M/10^8\msun$
(e.g. King \& Pounds, 2015, eqn (19), or King, 2023, eqn (6.17)). 

In the picture of the SMBH scaling relations derived by King (2003; 2005) and
summarized in King \& Pounds (2015) and King (2023), the $M - \sigma$ relation is a fundamental scaling for galaxies -- it is a direct result of momentum--driven (and later, energy--driven) black hole feedback, which gives $M = M_{\sigma} \propto \sigma^4$ with no free parameter. The linear relation with the central stellar bulge mass $M_b$ observed at low redshift comes about because that quantity is also $\propto \sigma^4$ (the Faber--Jackson relation), but for a different reason: momentum--driven feedback from stellar winds and supernovae. But this connection between $M$ and $M_b$ is acausal, i.e. there is no physical effect connecting the black hole and stellar mass and making them proportional to each other, so the $M/M_b$ ratio is not the fundamental scaling obeyed by supermassive black holes and their hosts.

Importantly, observations at $4 < z <7$ (Pacucci et al., 2023; Maiolino et al., 2023) 
show that the ratio of SMBH mass $M$ to stellar mass $M_b$ is noticeably larger than its typical value $\sim 10^{-3}$ at low redshift. The theoretical derivation of this ratio by Power et al. (2011) gives
\be 
\frac{M_{\sigma}}{M_b} \gtrsim \frac{1.25\times 10^{-3}h(z)}{\eta_c}\biggl[1 +
\frac{0.41\sigma_{200}}{h(z)}\biggr],
\ee
(where $\eta_c \sim1$ is an efficiency factor and $h(z) = H(z)/100\ {\rm km\ s^{-1}\ Mpc^{-1}} $ is the reduced Hubble parameter) and already suggests that the ratio should be at least $10\times$ larger at high redshift, because matter is virialized out to smaller radii at higher $z$. The mechanical feedback defining $M_{\sigma}$ may also reduce $M_{b}$, raising the ratio $M_{\sigma}/M_b$ still further.

\section*{DATA AVAILABILITY}
No new data were generated or analysed in support of this research.

\section*{ACKNOWLEDGMENTS}
I thank Ryan Hickox for a very helpful discussion, and the referee for a perceptive report that improved the paper.


{}

\section*{APPENDIX: Compton depth of the outflow}

The treatment here follows Shakura \& Sunyaev (1973) in noting that a black hole supplied with mass at a rate $\dot M$ significantly above the Eddington value
\be
\dot M_{\rm Edd} = \frac{L_{\rm Edd}}{\eta c^2} = \frac{4\pi GMc}{\eta c\kappa}
\ee
limits the mass rate actually accreting to the hole to a value $\simeq \dot M_{\rm Edd}$, and the total accretion luminosity to 
\be
L_{\rm Edd}(1 + \ln\dot m)
\ee 
(where 
$\dot m = \dot M/\dot M_{\rm Edd}$)
by expelling the excess matter  from each radius $R$ of the accretion disc in a quasispherical wind, with initial radial velocity $v(R)$ and total mass flow rate $\sim \dot M_{\rm Edd}$. Here $\eta c^2$ (with $\eta < 1$) is the binding energy per unit mass at the inner edge of the disc, $L_{\rm Edd}$ is the Eddington luminosity, and $\kappa$ the electron scattering opacity. The density of the wind at spherical radius $R$ is
\be 
\rho(R) \simeq \frac{\dot M_E}{4\pi R^2v(R)} = \frac{GM}{\eta c \kappa R^2 v(R)} > \frac{GM}{\eta\kappa c^2R^2}
\ee
since $v(R) < c$,
so the electron scattering optical depth through the wind from the gravitational radius $R_g > GM/c^2$ to infinity is
\be
\tau = \int_{R_g}^{\infty} \kappa\rho(R){\rm d}R > \frac{GM}{\eta c^2 R_g} = {1\over \eta} > 1.
 \ee
Accordingly the outflow is Compton thick in the radial direction, and only velocities from matter above the photosphere are in principle measurable.
\end{document}